# Orbital Angular Momentum Generating Thin-Film Conformal Metasurface for Backscattering Control

Yury M. Meleshin,[1,*] Maxim V. Azarov,[1] Artem A. Airapetian,[1] Konstantin S. Lyalin[1]

[1]Institute of Microdevices and Control Systems of the National Research University of Electronic Technology, Moscow, 124498, Russia

This paper presents the design, simulation, and experimental verification of a flexible thin-film conformal metasurface for backscattering control via orbital angular momentum (OAM) wave generation in the X-band (9.5–10.5 GHz). The metasurface comprises asymmetric square loop meta-atoms fabricated on an air-equivalent dielectric substrate ($\varepsilon_r \approx 1$) with subwavelength thickness ($< \lambda/15$ at 10 GHz). Analytical modeling define the phase difference requirement ($\arg(r_{yy}) - \arg(r_{xx}) \approx 180° \pm 25°$) for spin-to-orbital angular momentum conversion, enabling OAM mode $l = +1$ generation upon circularly polarized wave reflection. Full-wave EM simulations and antenna array theory predictions confirmed a characteristic radiation pattern null along the OAM propagation axis. Experimental prototypes were fabricated using adhesive-backed foil on foam-core substrates, demonstrating: 4 dB reflected power reduction for the OAM-generating metasurface compared to a uniform-phase reference; 7 dB reflected waves reduction versus solid metal including for various radii of rounding surfaces. The measured suppression bandwidth reached 1 GHz, though alignment sensitivity due to the narrow null zone was identified as a limitation. Additional radial phase gradients for beam defocusing reduced performance by 1–2 dB due to main lobe distortion. Discrepancies between analytical models and measurements underscored the importance of mutual coupling effects in complex phase distributions. This work confirms thin-film conformal metasurfaces as a viable solution for controllable backscattering.

## I. INTRODUCTION

Active research into quasi-optical wave beams carrying orbital angular momentum (OAM) in the radio frequency range is driven by the potential applications using the unique properties of such waves in radio communications and radar systems. However, several characteristics of OAM waves, also known as vortex waves, pose challenges for their successful implementation. Consequently, their practical use is currently limited to laboratory settings and a small number of field experiments. Nevertheless, the very properties of OAM waves that have a detrimental effect in some applications can be utilized beneficially in others. For instance, their characteristic amplitude and phase distribution can be applied in areas requiring control of reflectance.

OAM waves can be generated using various methods, one of which involves the use of metasurfaces, particularly reflective types [1]. Metasurfaces are artificial structures that enable manipulation of amplitude, phase, and polarization at the subwavelength level. The extensive scientific foundation in this field, accumulated through decades of research on OAM waves, offers a wide selection of reflective metasurfaces. These metasurfaces generate various OAM modes across different frequency ranges for both linearly and circularly polarized incident waves [2], [3], [4], [5], [6], [7], [8], [9], [10], [11], [12].

Nevertheless, research into the application of reflective metasurfaces for reducing reflectance (backscattering control) has only begun in the last few years and remains an active area of investigation [13], [14], [15], [16], [17]. The application of OAM waves, alongside other phase distributions, provides a flexible means to control the backscattering profile of the metasurface.

This paper is structured as follows: Section 2 presents the problem of backscattering control, a brief overview of main approaches, and the rationale for the chosen research direction. Section 3 provides the theoretical background required for the analytical calculation of the metasurface, along with data and simulation results for various metasurface configurations. Section 4 describes the experimental setup and presents the results of experiments conducted with the proposed metasurfaces. The work specifically addresses the case of backscattering upon normal incidence reflection of a plane wave from a flat surface, investigating both the shape and intensity of the entire radiation pattern.

## II. Backscattering Control Using Metasurfaces

### A. The problem of backscattering control

The control, specifically the reduction of backscattering within a specified frequency range, is a relevant task for both radio communication and radar applications. It enables the minimization of spurious scattering in complex electromagnetic environments. In some cases, the ability to control and to shape the required backscattering profile is a fundamental aspect of the problem solved by radio engineering systems. Typically, this is achieved through the optimization of object shapes, the use of radar-absorbing materials (RAM), and scattering structures. The mentioned traditional RAM are often composites with specific dielectric and magnetic properties. Within these materials, the energy of incident electromagnetic waves is converted into heat. To achieve this, they are implemented as complex, sometimes multilayer structures, resulting in significant thickness [4], [5]. Consequently, their application often degrades performance characteristics and imposes operational limitations.

## B. Application of metasurfaces

Engineered scattering meta-structures alter the backscattering profile according to a desired pattern through controlled manipulation of the phase and amplitude of the reflected signal. They can be fabricated from readily available and simple materials. Advances in metamaterials enable their use across wide frequency ranges for various polarizations. Metasurfaces, featuring subwavelength thickness, potentially can be deployed on complex curved surfaces without degrading aerodynamic or other performance characteristics of objects.

Two main approaches can be identified for designing metasurfaces for backscattering control: gradient metasurfaces, which impart an additional phase variation along the surface to emulate diffuse reflection; coding metasurfaces [13], [14], [15] [18], [19], [20], where the impedance varies according to a specified pattern to create pseudorandom or other specialized scattering patterns. The latter category includes metasurfaces generating OAM waves, capable of producing OAM waves of various modes across a broad frequency range [6], [7], [8], [12], [24], [25] for different polarizations [3], [10], [11]. A schematic representation of the metasurface that forms OAM waves when reflecting a normal incident plane (irrespective of the polarization) wave is shown in Figure 1.

The operating principle for OAM wave generation in most metasurfaces involves the controlled phase modulation of the reflected signal with minimal amplitude variation. This relies on the fundamental unit cell of the metasurface – the meta-atom – a structure comprising one or more elements arranged on single or multiple dielectric layers. A meta-atom enables phase control spanning 0 to 360 degrees, achieved either by varying its geometry or through rotation around its axis, using the geometric phase principle (Pancharatnam-Berry phase) [21], [22], [23].

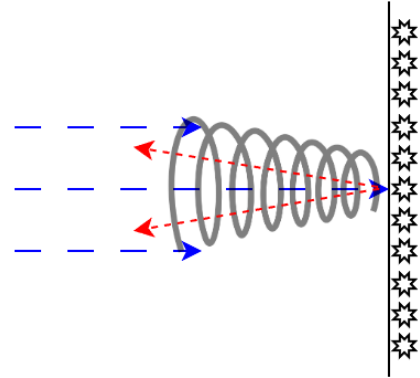

FIG. 1. Scattering of an incident wave onto a flat surface using a metasurface generating OAM waves.

Research teams have published results in recent years on metasurface and OAM wave applications for controlled scattering [15], [16], [17], [26], [27]. The diversity of achieved results allows for assessment of the approach's prospects, prompting the task of developing such a metasurface on a flexible thin dielectric substrate for operation in the X-band frequency range with a 1 GHz bandwidth.

## III. Thin-Film Metasurface for Generating OAM Waves

### A. Analytical calculation of a meta-atom

Given the long-standing and widespread use of metamaterials in high-frequency engineering [28], various authors have performed calculations and experimental validations of design parameters for typical structures used in metamaterial development [29], [30], [31]. Leveraging these results, it is appropriate to reduce the metasurface design problem with specified properties to determining the specific meta-atom parameters control its impedance. In this case, the air-metasurface reflective system can be represented as show in Figure 2.

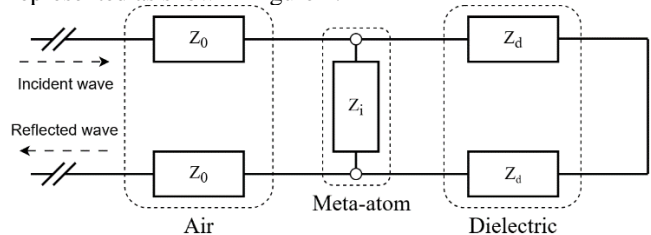

FIG. 2. Equivalent scheme of reflective metasurface.

Considering the necessary conditions for generating OAM waves upon reflection of an incident wave and the representation equations presented in [25], [32], [33], [34], the reflection coefficients for spin-to-orbital angular momentum conversion can be expressed as follows:

$$r_{ll} = 0.5 * \left((r_{xx} - r_{yy}) + j(r_{xy} + r_{yx})\right)e^{-2jk\varphi}; \quad (1)$$
$$r_{lr} = 0.5 * \left((r_{xx} + r_{yy}) + j(r_{yx} - r_{xy})\right); \quad (2)$$
$$r_{rl} = 0.5 * \left((r_{xx} + r_{yy}) - j(r_{yx} - r_{xy})\right); \quad (3)$$
$$r_{rr} = 0.5 * \left((r_{xx} - r_{yy}) - j(r_{xy} + r_{yx})\right)e^{2jk\varphi}, \quad (4)$$

where $r_{ll}, r_{rr}$ – co-polarized reflection coefficients (circular basis), $r_{lr}, r_{rl}$ – cross-polarized reflection coefficients (circular basis), $r_{xx}, r_{yy}$ and $r_{xy}, r_{yx}$ – corresponding coefficients for linear polarization.

It then becomes possible to formulate requirements accounting for phase control of the reflected wave through meta-atom rotation and its symmetry, as also presented in the aforementioned works:

$$r_{yy} + r_{xx} = 0, \ r_{yy} - r_{xx} = 2, \quad (5)$$

in other words, it can be expressed as:

$$|r_{yy}| = |r_{xx}| = 1, \quad arg(r_{yy}) - arg(r_{xx}) = \pm\pi.$$

For analytical convenience, the requirement of a $\pi$ phase difference between reflection coefficients across the frequency band can be expressed as:

$$\frac{d\phi}{d\omega} \approx 0, \phi(\omega_1) = \phi(\omega_2) = \pi, \quad (6)$$

where $\omega_1, \omega_2$ – the boundaries of the frequency range.

Satisfying these meta-atom requirements across the specified frequency band ensures spin-to-orbital angular momentum conversion within that frequency range. Meeting these requirements over a broad bandwidth is achievable using parallel LC-resonator structures configured as rectangular loops (Fig. 3), as demonstrated in the mentioned works [29], [31].

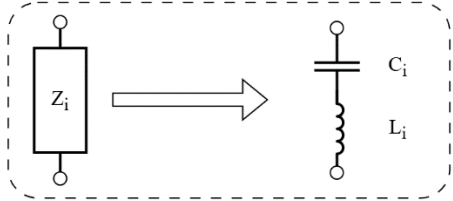

Fig. 3. Equivalent circuit of a meta-atom.

Returning to the equivalent circuit and reflection coefficient requirements, the following can be expressed in terms of impedance:

$$r_{xx} = |r_{xx}|e^{j\varphi_x} = \frac{Z_i^{in} - Z_0}{Z_i^{in} + Z_0}, (i = 1); \quad (7)$$

$$r_{yy} = |r_{yy}|e^{j\varphi_y} = \frac{Z_i^{in} - Z_0}{Z_i^{in} + Z_0}, (i = 2), \quad (8)$$

where the correlation between impedance and equivalent capacitances/inductances is given by:

$$Z_i^{in} = \frac{jZ_d \tan(\beta h) * \left(j\omega L_i + \frac{1}{j\omega C_i}\right)}{jZ_d \tan(\beta h) + j\omega L_i + \frac{1}{j\omega C_i}}. \quad (9)$$

For each linear polarization component, the impedance $Z_i^{in}$ is calculated with corresponding index i. For simplicity of analysis and computations, this need only be done for two orthogonal directions, corresponding to the previously used reflection coefficients $r_{xx}$ and $r_{yy}$, assigned indices i=1 and i=2 respectively are shown in Figure 4.

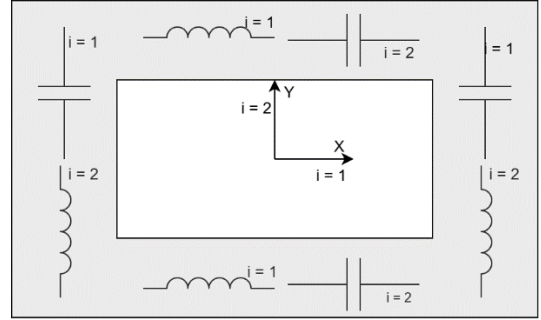

FIG. 4. Schematic representation of a meta-atom based on a square ring with equivalent parameters for different polarizations.

Based on the metasurface requirements for a center frequency of 10 GHz and 1 GHz operational bandwidth, the following meta-atom capacitance and inductance values were calculated and selected: $L_1 = 2.38$ nH, $L_2 = 1.25$ nH, $C_1 = 0.075$ pF, $C_2 = 0.040$ pF. The corresponding reflection coefficients are shown in Figure 5. These electrical parameters ensure a 180° phase difference with ±25° tolerance within the 9–11 GHz frequency range.

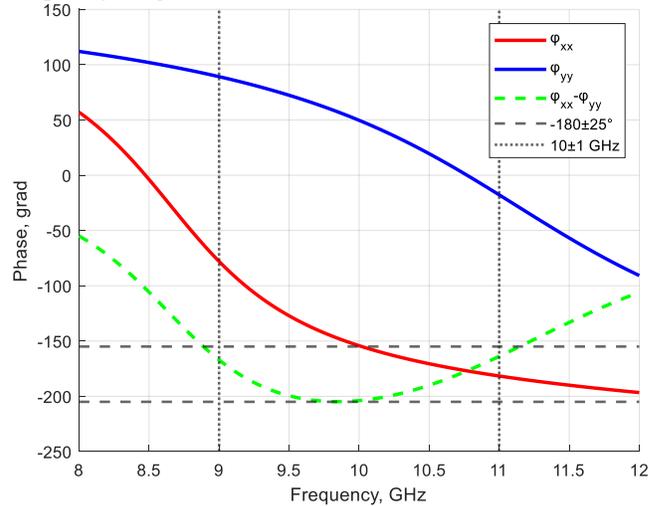

FIG. 5. Phase graphs of calculated reflection coefficients $r_{yy}, r_{xx}$ and their differences in the frequency range from 8 to 12 GHz.

The analytical modeling yields the following conclusions:
- Inductance L1 is inversely proportional to the phase difference in the upper frequency range;
- Inductance L2 is directly proportional to the phase difference in the lower frequency range;
- Capacitance C1 is inversely proportional to the

phase difference in the mid-to-high frequency ranges;
- Capacitance C2 is directly proportional to the phase difference in the lower frequency range.

Next, the relationship between the electrical characteristics and geometric parameters of the meta-atom must be established. For this purpose, we refer to equations from [29], [35] to derive the governing dependencies for a symmetric square loop structure:

$$L = \frac{Z_0}{\omega} \frac{d}{p} F(p, 2s, \lambda); \qquad (10)$$

$$C = \frac{Y_0}{\omega \varepsilon_{eff}} \frac{4 d \varepsilon_{eff}}{p} F(p, g, \lambda), \qquad (11)$$

where p – unit cell period (loop spacing), s – loop line width, g – loop gap, d – Outer dimension of square loop.

Here, the functions $F(p, 2s, \lambda)$ and $F(p, g, \lambda)$ for TE and TM polarizations are defined as:

$$F(p, 2s, \lambda)_{TE} = \frac{p}{\lambda} \cos\theta \left( \ln\left(\csc\frac{\pi s}{p}\right) + G(p, 2s, \lambda, \theta) \right); \quad (12)$$

$$F(p, g, \lambda)_{TE} = \frac{p}{\lambda} \sec\theta \left( \ln\left(\csc\frac{\pi g}{2p}\right) + G(p, g, \lambda, \theta) \right); \quad (13)$$

$$F(p, 2s, \lambda)_{TM} = \frac{p}{\lambda} \sec\phi \left( \ln\left(\csc\frac{\pi s}{p}\right) + G(p, 2s, \lambda, \phi) \right); \quad (14)$$

$$F(p, g, \lambda)_{TM} = \frac{p}{\lambda} \cos\phi \left( \ln\left(\csc\frac{\pi g}{2p}\right) + G(p, g, \lambda, \phi) \right). \quad (15)$$

In turn, the functions $G(p, 2s, \lambda, \theta)$ and $G(p, g, \lambda, \phi)$ are correction terms and can be expressed as follows:

$$G\left(p, \frac{2s}{g}, \lambda, \frac{\theta}{\phi}\right) = \frac{A}{B}, \qquad (16)$$

where

$$A = 0{,}5(1 - \beta^2)^2 \left( \left(1 - \frac{\beta^2}{4}\right)(C_{+1} + C_{-1}) + 4\beta^2 C_{+1} C_{-1} \right),$$

$$B = \left(1 - \frac{\beta^2}{4}\right) + \beta^2 \left(1 + \frac{\beta^2}{2} - \frac{\beta^4}{8}\right)(C_{+1} + C_{-1}) + 2\beta^6 C_{+1} C_{-1}, \qquad (17)$$

where $\beta = \sin(\pi s / 2p)$, and the coefficients $C_{\pm 1}$ for TE and TM polarizations are:

$$C_{\pm 1}^{TE} = \frac{1}{\sqrt{\left(\frac{p \sin\theta}{\lambda} \pm 1\right)^2 - \frac{p^2}{\lambda^2}}} - 1, \qquad (18)$$

$$C_{\pm 1}^{TM} = \frac{1}{\sqrt{1 - \left(\frac{p \cos\phi}{\lambda}\right)^2}} - 1. \qquad (19)$$

Using (10) – (19) equations, the geometric parameters of the square loop can be determined based on the required electrical characteristics. We introduce additional geometric designations for the meta-atom: $d1, d2, s1, s2$ illustrated in Figure 6.

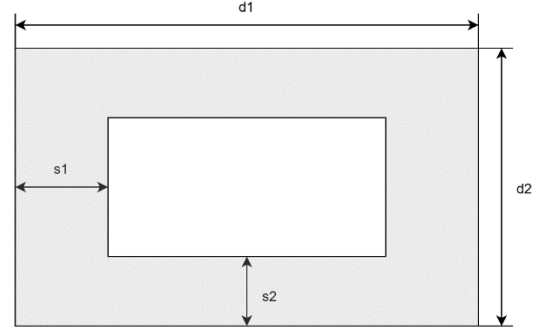

FIG. 6. The legend of dimensions by meta-atoms.

Applying the equations to achieve the target electrical parameters ($L_1 = 2.38$ nH, $L_2 = 1.25$ nH, $C_1 = 0.075$ pF, $C_2 = 0.040$ pF), the following meta-atom dimensions are obtained: $d_1 = 13.77$ mm, $s_1 = 2.93$ mm, $d_2 = 11.56$ mm, $s_2 = 4.87$ mm.

Mapping geometric parameters to electrical characteristics yields the following dependencies:
- Dimension d1 primarily affects capacitance C1 (x-direction), with dominant influence in the low-frequency;
- Dimension s1 primarily affects inductance L1 (x-direction), with dominant influence in the high-frequency;
- Dimension d2 primarily affects capacitance C2 (y-direction), with dominant influence in the low-frequency;
- Dimension s2 primarily affects inductance L2 (y-direction), with dominant influence in the high-frequency.

### B. Meta-atom modeling

The analytical meta-atom calculation yielded initials geometric parameters, which served as starting values for EM simulation in a computational electromagnetics environment. A frequency-domain solver was used to model a single meta-atom, with unit cell boundary conditions in the planar directions (x and y) and Floquet ports (z-direction) for the first two modes.

The initial values required iterative adjustment to meet specifications based on simulation results. The finalized parameters are presented in Table 1:

TABLE I. Values of geometric parameters of the meta-atom.

| Parameter | Calculated value, mm | Simulated value, mm | Difference |
|---|---|---|---|
| d1 | 13,77 | 12,05 | -10% |
| s1 | 2,93 | 5,6 | +91% |
| d2 | 11,56 | 9,35 | -19% |
| s2 | 4,87 | 1,87 | +160% |

The Figure 7 shows the final shape and dimensions of the resulting meta-atom.

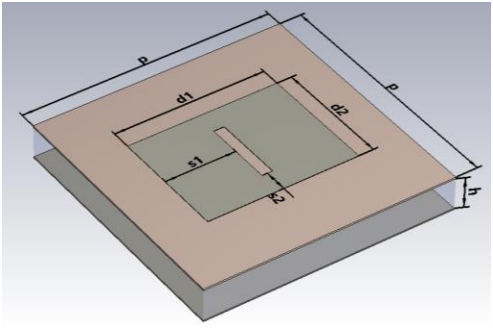

FIG. 7. The 3D view of a meta-atom obtained from the simulation results; p = 20 mm is the size of one meta-atom, the repetition period; h = 2 mm is the height of the air-filled layer; d1, d2, s1, s2 are the corresponding calculated and simulated dimensions of the meta-atom structure.

The simulation was carried out in the frequency range from 8 GHz to 12 GHz, the results for the reflection coefficients of linear polarizations are shown in the Figure 8, which shows the amplitudes, phases and phase difference.

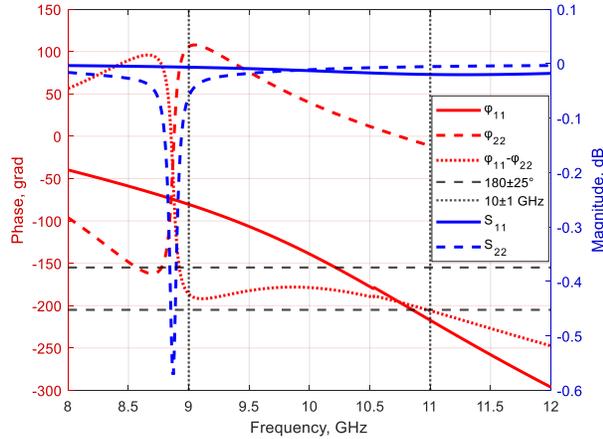

FIG. 8. Graphs of simulated reflection coefficients for linear polarizations and their phase differences.

The graphs show that the reflection coefficients in the frequency range from 9 GHz to 11 GHz are close to zero dB in amplitude, and their phase difference remains on the order of 180° with a tolerance of ±25° which was required. Since the meta-atom will convert the spin angular momentum into an orbital angular momentum, it is also interesting to look at the reflection coefficients for modes with circular polarization are shown in the Figure 9.

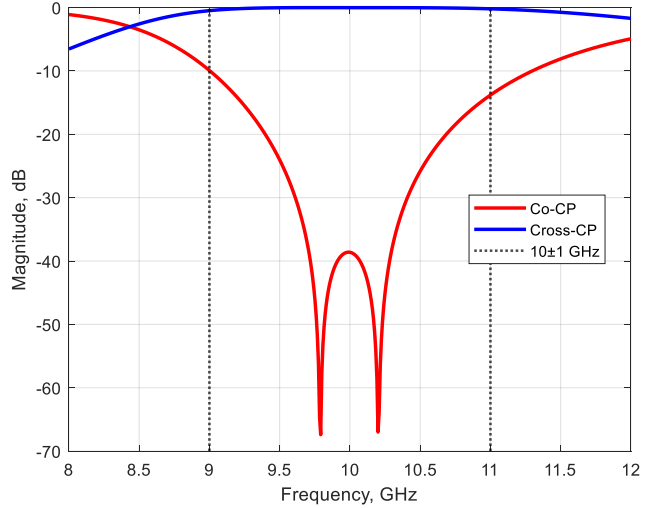

FIG. 9. Simulated reflection coefficients for circular polarization

It can be seen from the graph that the condition for converting circularly polarized waves into OAM waves is fulfilled when the reflection coefficient (co-component) takes minimum values, and the polarization conversion coefficient (cross component) takes maximum values.

### C. Metasurface modeling

Generating the target OAM mode requires implementing a specific phase distribution across the metasurface. This distribution is described by the equation:

$$\Phi_{OAM} = l\tan^{-1}(y_i/x_i), \qquad (20)$$

where l is the OAM mode index, and $y_i$, $x_i$ denote the element coordinates.

Thus, the required phase distributions for generating OAM modes ±1 on a 9×9 element metasurface are as follows on Figure 10:

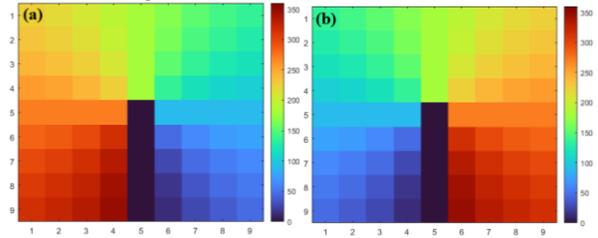

FIG. 10. Calculated phase distributions for the formation of OAM modes +1 (a) and -1 (b).

Given the radiation pattern of a single element (Fig. 11) and the designed phase distributions, the overall metasurface radiation pattern can be evaluated using antenna array theory principles are as follows on Figure 12.

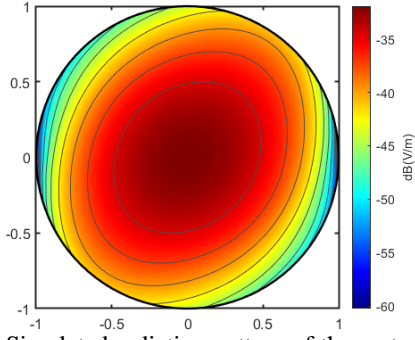

FIG. 11. Simulated radiation pattern of the meta-atom at a frequency of 10 GHz, abs.

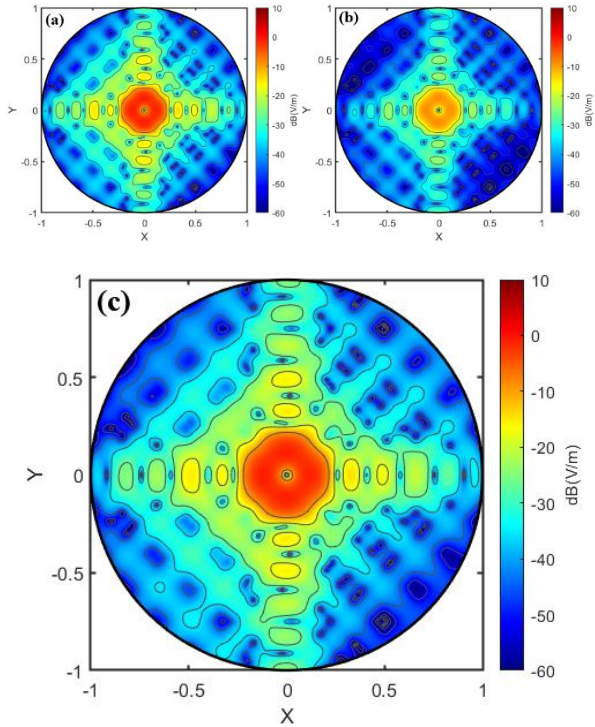

FIG. 12. Calculated directional pattern of a 9-by-9 metasurface generating an OAM mode +1, 10 GHz, co-CP (a), cross-CP (b), abs (c).

The radiation pattern exhibits a characteristic deep null along the propagation axis of OAM waves. Thus, preliminary modeling confirms the viability of the chosen approach, warranting progression to full metasurface simulation.

To achieve the required phase distribution, meta-atoms are rotated by half the target phase according to the Pancharatnam-Berry geometric phase principle, yielding the rotation angle for each element (Fig. 13).

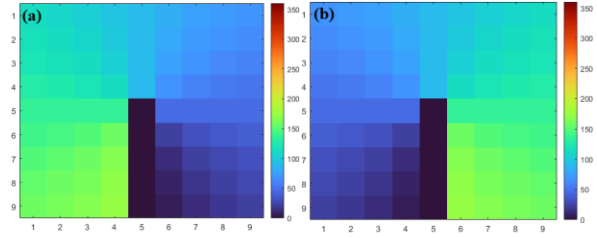

FIG. 13. Calculated values of the meta-atom rotations for the formation of OAM modes are +1 (left) and -1 (right).

Simulations employ a time-domain solver with open boundaries on all sides. Results provide far-field patterns at specified frequencies. The metasurface model integrates individual meta-atoms with predetermined rotations, maintaining the unit cell periodicity from standalone simulations. The 9×9 array (81 elements total) spans 180 mm × 180 mm. Illumination is provided by a circularly polarized plane wave. The radiation pattern at 10 GHz is shown below on Figure 14.

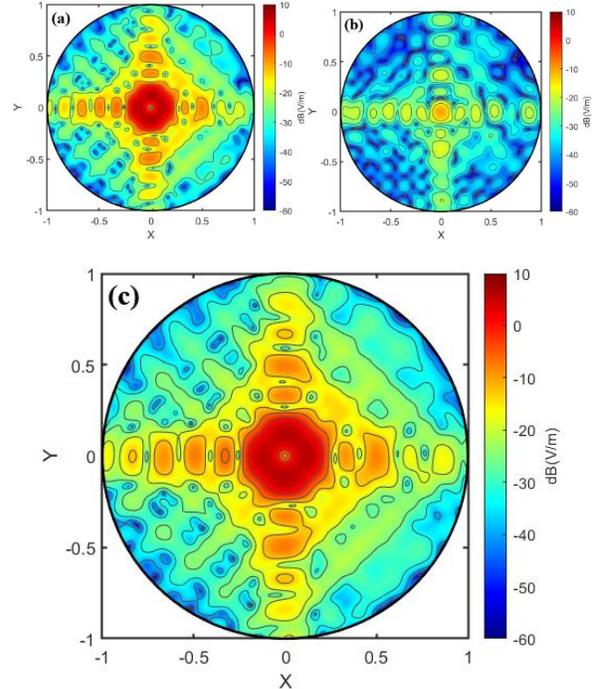

FIG. 14. Simulated directional pattern of a 9-by-9 metasurface generating an OAM mode +1, 10 GHz, co-CP (a), cross-CP (b), abs (c).

It should be noted that the full-wave EM simulation results for the metasurface qualitatively match the analytically derived radiation pattern in terms of total electric field amplitude. However, differences exist between polarization components. Specifically, the EM simulations do not demonstrate OAM wave formation in the cross-polarized component of the reflected wave, whereas the analytical model predicted OAM generation for both polarization components.

### D. *Additional phase distributions*

Concurrently with the phase distribution for generating a specified OAM mode, additional phase distributions can be utilized for backscattering control. Such distributions may include linear gradients, radial gradients, or coding patterns to create multi-beam radiation patterns. This work examines supplementary radial distributions.

### E. *Radial* gradient phase distributions

The radial gradient distribution follows the expression:
$$\Phi_r = \frac{2\pi}{\lambda_0}\left(\sqrt{X_i^2 + Y_i^2 + F^2} - F\right). \quad (21)$$

This distribution enables additional focusing/defocusing of the main beam lobe for more flexible pattern shaping. The radial gradient phase distributions for focal lengths of 400, 700, and 1100, along with the resulting phase profiles (sum of gradient phases and OAM mode +1 distribution), are presented in Figures 15-17.

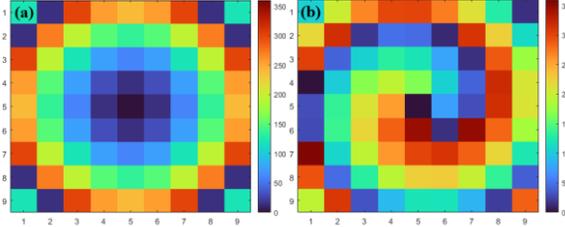

FIG. 15. Radial phase distribution for F = 400 (a) and the final phase distribution (b), which is the sum of the radial and OAM mode +1.

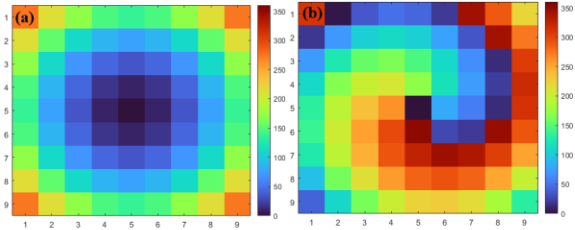

FIG. 16. Radial phase distribution for F = 700 (a) and the final phase distribution (b), which is the sum of the radial and OAM mode +1.

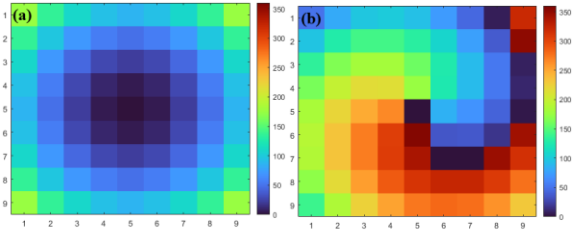

FIG. 17. Radial phase distribution for F = 1100 (a) and the final phase distribution (b), which is the sum of the radial and OAM mode +1.

The calculation and simulation results of OAM metasurfaces with an additional radial phase distribution are shown in the Figures 18-20.

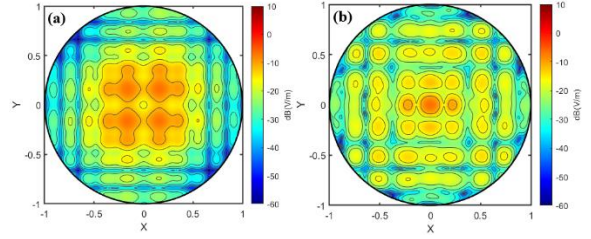

FIG. 18. Radiation patterns of an OAM metasurface with an additional radial phase distribution at F = 400, calculated on (a), simulated on (b).

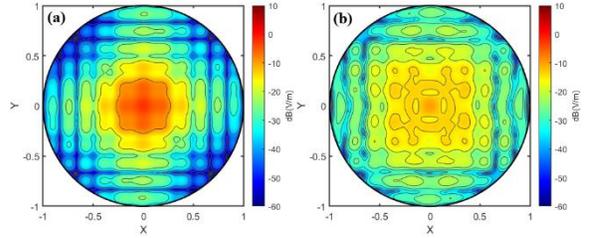

FIG. 19. Radiation patterns of an OAM metasurface with an additional radial phase distribution at F = 700, calculated on (a), simulated on (b).

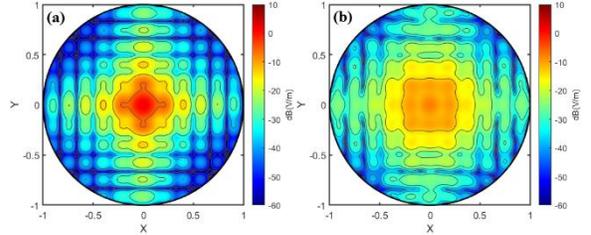

FIG. 20. Radiation patterns of an OAM metasurface with an additional radial phase distribution at F = 1100, calculated on (a), simulated on (b).

The figures reveal significant differences not only in numerical values but also in the resulting radiation pattern shapes. For clearer comparison, co- and cross-polarized components of reflected waves for the F = 700 case are shown on Figures 21 and 22, contrasting different computational approaches.

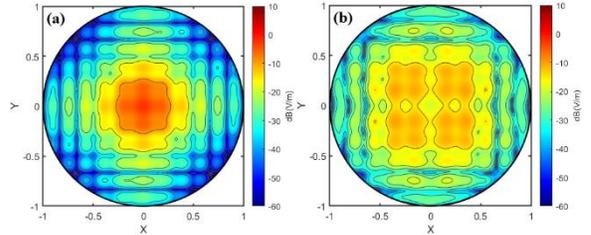

FIG. 21. Radiation patterns at F = 700 for the polarization co-component, calculated on (a), modeled on (b).

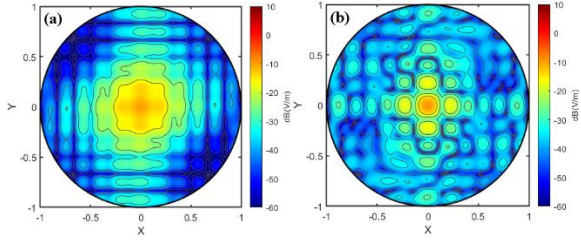

FIG. 22. Radiation patterns at F = 700 for the polarization cross-component, calculated on (a), modeled on (b).

These results demonstrate differences in both co- and cross-polarized components, resulting in fundamentally different resultant radiation patterns. In this case, preliminary analytical modeling using antenna array theory fails to yield satisfactory predictions of the metasurface's radiation behavior.

Full-wave EM simulations confirm transformed radiation patterns where:
- The characteristic deep null along the OAM propagation axis disappears;
- The main beam lobe splits into multiple lobes;
- Peak gain decreases by 10-20 dB depending on focal length.

## IV. Experiment

To validate the proposed solutions, experiments were conducted with the previously described OAM-generating metasurfaces. In addition to the described checks, an experiment was conducted with different bending radii of the metasurface OAM to determine the operability. For comparison, the following configurations were selected: Reference metasurface: Uniform array (no phase gradient) shown in Figure 23(b), OAM mode +1 metasurface: Shown in Figure 23(c), OAM mode +1 with supplementary gradient: Shown in Figure 23(d).

Samples were fabricated using adhesive-backed foil paper and foam-core foil sheets. Metasurface patterns were etched on paper-backed foil, while foam-core material ($\varepsilon_r \approx 1$) served as an air-equivalent dielectric spacer. A solid metal reference sample was also manufactured. Test specimens are shown in Figure 23.

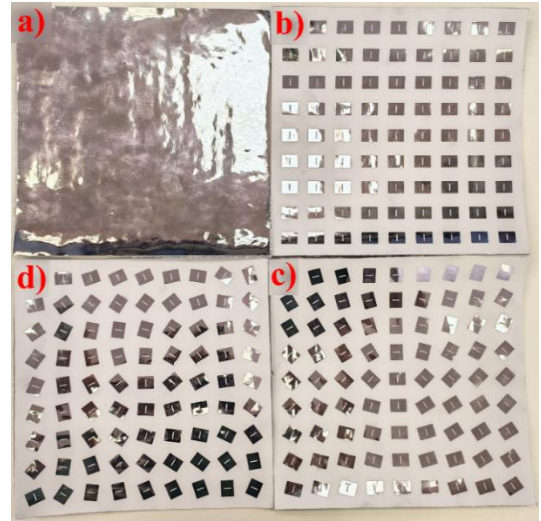

FIG. 23. (a) a solid metallized sheet; (b) an equivalent metasurface; (c) a metasurface for generating the OAM mode +1; (d) a metasurface for generating the mode +1 with an additional radial phase distribution.

The experimental setup is illustrated in Figure 24. Reflected power was measured using a vector network analyzer (VNA) connected to two antennas, recording transmission coefficients ($S_{21}$) as shown on Fig. 25. The transmitting antenna provided circular polarization, while the receiving antenna had linear polarization. To obtain absolute power values the receive antenna was rotated to measure horizontal and vertical polarization components when data was combined into absolute reflected power values. Baseline transmission between antennas was maintained below -70 dB using radio-absorbing materials and spatial separation. The distance from the antennas to the metasurface was 100 centimeters. Measurements were then performed with metasurface samples and solid metal sheet reference to quantify reflected power levels.

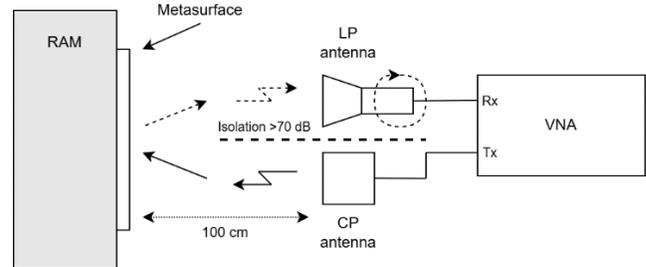

FIG. 24. The scheme of the experiment.

Figure 26 shows metasurfaces mounted on various rigs that provide the required bending radius.

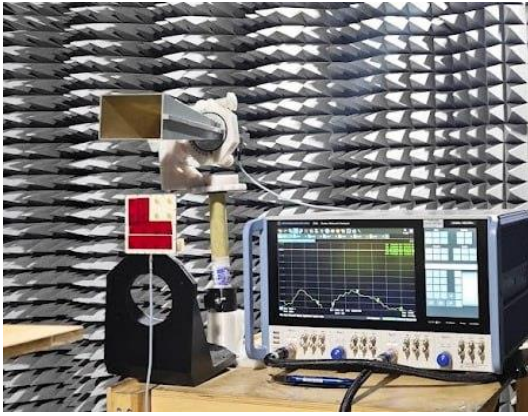

FIG. 25. Picture of a measuring stand assembled to measure the transmission coefficient using VNA.

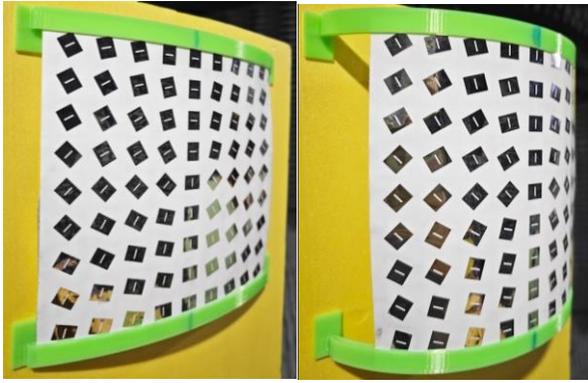

FIG. 26. Two installation options for the metasurface: left – with small bend radius, right – with large bend radius.

Figures 27-29 presents normalized transmission coefficient values at three frequencies (Fig. 27 for 9.5 GHz, Fig. 28 for 10 GHz, Fig. 29 for 10.5 GHz) for the three metasurfaces and solid metal sheet, normalized to the equivalent metasurface.

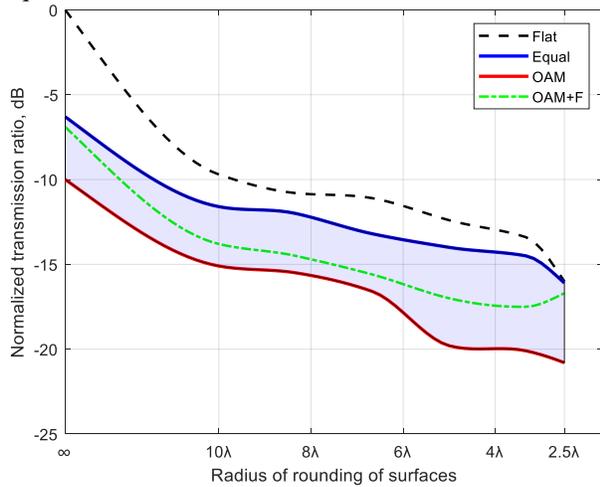

FIG. 27. Normalized transmission coefficients at frequency 9.5 GHz versus surface bending radius, ranging from a flat configuration to a bend radius of 2.5 wavelengths.

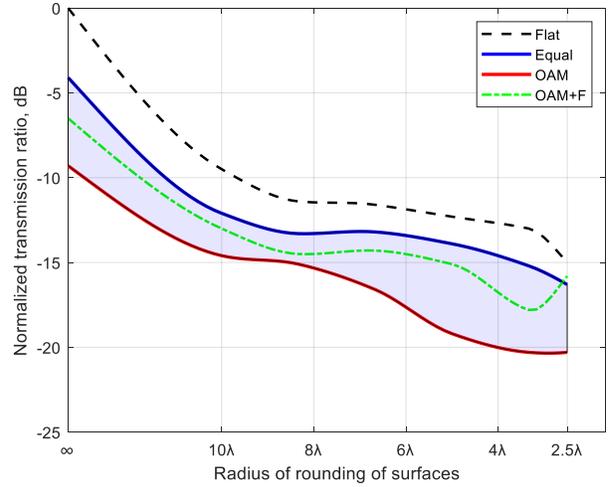

FIG. 28. Normalized transmission coefficients at frequency 10 GHz versus surface bending radius, ranging from a flat configuration to a bend radius of 2.5 wavelengths.

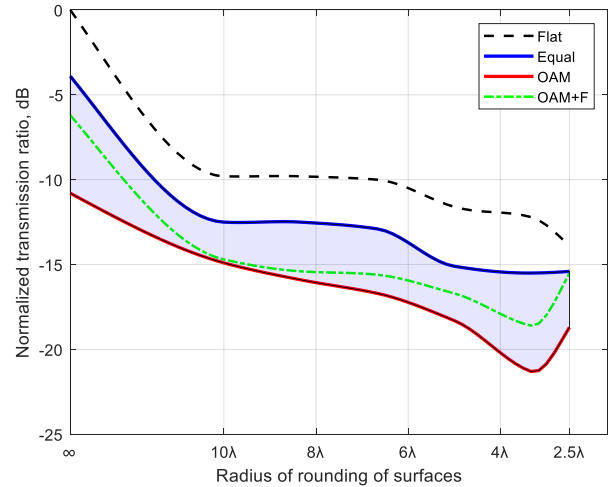

FIG. 29. Normalized transmission coefficients at frequency 10.5 GHz versus surface bending radius, ranging from a flat configuration to a bend radius of 2.5 wavelengths.

TABLE II. Comparison between the performances of the proposed metasurface and previous works.

| Ref./year | Frequency range, GHz | Metasurface type | Reduction Backscattering /RCS (polarization) | Thickness (in λ at central freq.) | Flexible |
|---|---|---|---|---|---|
| [18]/2025 | 13–20 | Transparent PB-coded ITO on PET | >10 dB (LP) | 0.10λ | Yes |
| [7]/2024 | 95–105 | Reflective PB-coded | — (LP) | 0.08λ | No |
| [15]/2022 | 5–11 | PB-coded OAM | ≈ 9 dB (LP) | 0.08λ | No |
| [21]/2021 | 17 | Reflective PB-coding | — (CP) | 0.11λ | No |
| [22]/2020 | 11–17 | Single-layer PB-phase reflection-type | — (CP) | 0.14λ | No |
| [19]/2023 | 6.9–14.5 | PB-coding reflective polarizer (FR4) | ≈ 5 dB (LP) | 0.11λ | No |
| [20]/2023 | 10.8–31.3 | Integrated polarization conversion + diffusion + absorption; PB-coded | ≥10 dB (simulation only) (LP) | 0.14λ | Yes |
| **This work** | 9.5–10.5 (meas.) 9–11 (sim.) | **Thin-film OAM-generating** | ≈7 dB (vs metal) ≈4 dB (vs metasurface) | **0.07λ** | **Yes** |

Results indicate:

The equivalent metasurface reduces transmission ratio by 3 dB for all values of bends relative to solid metal solely due to meta-atom polarization conversion properties.

The OAM mode +1 metasurface provides an additional 4 dB transmission reduction across the band and values of bends, attributable to OAM wave characteristics.

The metasurface with supplementary gradient phase distribution performs 2-3 dB worse than the OAM-only structure.

Differences between experimental and simulated results stem from the narrow null in the radiation pattern of OAM mode +1 metasurfaces. Minor deviations from the propagation axis degrade backscattering reduction measurements along the surface normal. Comprehensive validation requires experimental characterization of the reflected field's amplitude-phase distribution in the transverse plane.

Nevertheless, the experiment demonstrated that these metasurfaces can be utilized for backscattering control even when deployed on curved surfaces.

**V. Conclusion**

This comprehensive study confirms the promise of reflective metasurfaces generating orbital angular momentum (OAM) waves for effective backscattering control in the X-band with operating 1 GHz band (9.5–10.5 GHz). The key achievement is the development and experimental verification of a flexible thin-film conformal metasurface with subwavelength thickness (< λ/15 at 10 GHz), based on asymmetric square ring meta-atoms. Analytical calculations and full-wave simulations demonstrate that the structure enables spin-to-orbital angular momentum conversion when satisfying the reflection coefficient phase difference condition for orthogonal linear polarizations: $\arg(r_{yy}) - \arg(r_{xx}) \approx 180° \pm 25°$ across the 1 GHz target bandwidth. This ensures efficient generation of OAM mode $l = 1$ upon reflection of a circularly polarized wave.

Monostatic transmission coefficient measurements demonstrated backscattering reduction along the surface normal. The OAM mode +1 metasurface achieved 4 dB reflected power reduction in the 9.5–10.5 GHz range compared to the uniform-phase reference metasurface (and 7 dB reduction versus solid metal) including for various radii of rounding surfaces. This reduction directly correlates with the characteristic deep null along the OAM propagation axis predicted by simulations. However, adding a radial gradient phase distribution for beam defocusing degraded performance by 2-3 dB due to main lobe broadening and distortion. Discrepancies between analytical (array factor) and full-wave EM models highlight the critical importance of accounting for mutual coupling and polarization effects when implementing complex phase distributions.

Key technological aspects include accessible fabrication materials and a design methodology enabling precise meta-atom realization. The primary limitation is the narrow backscattering null in the radiation pattern, requiring precise system alignment and reducing operational reliability. Promising directions include higher-order OAM modes with broader nulls and hybrid coding-phase approaches. These results demonstrate the feasibility of ultrathin radar coatings with tunable reflectivity. Future work will focus on bandwidth enhancement, non-uniform/coded phase distributions, and stability analysis under substrate deformation.